\documentclass[pdflatex,sn-mathphys-num]{sn-jnl}% Math and Physical Sciences Numbered Reference Style
\usepackage{graphicx}%
\usepackage{multirow}%
\usepackage{amsmath,amssymb,amsfonts}%
\usepackage{amsthm}%
\usepackage{mathrsfs}%
\usepackage[title]{appendix}%
\usepackage{xcolor}%
\usepackage{textcomp}%
\usepackage{manyfoot}%
\usepackage{booktabs}%
\usepackage{algorithm}%
\usepackage{algorithmicx}%
\usepackage{algpseudocode}%
\usepackage{listings}%
\usepackage{graphicx}
\usepackage{dcolumn}
\usepackage{bm}
\usepackage{epstopdf}
\usepackage{algorithm}
\usepackage{algpseudocode}
\hypersetup{colorlinks=true, allcolors=blue}
\usepackage{amsmath,amsthm, amssymb, amsfonts}%
\usepackage{float}
\usepackage{epsfig}
\usepackage{mathrsfs}
\usepackage{multirow}
\usepackage[all]{xy}
\usepackage{pbox}
\usepackage{verbatim}
\usepackage{braket}
\usepackage{mathtools}
\usepackage{tikz}
\usepackage{xcolor}
\usepackage{xfrac}
\usepackage{cleveref}
\usepackage[caption=false]{subfig}
\usepackage{booktabs}
\usepackage{hyperref}

\theoremstyle{thmstyleone}%
\theoremstyle{thmstyletwo}%

\theoremstyle{thmstylethree}%

\newcommand{\oneQRp}{\left[\begin{matrix}e^{- i \omega t} & 0\\0 & e^{i \omega t}\end{matrix}\right]}
\newcommand{\oneQRm}{\left[\begin{matrix}e^{i \omega t} & 0\\0 & e^{- i \omega t}\end{matrix}\right]}
\newcommand{\oneQotocTwo}{2 - 2 \cos{\left(4 \omega t \right)}}
\newcommand{\oneQotocThree}{- 4 \cos^{2}{\left(4 \omega t \right)} + 8 \cos{\left(4 \omega t \right)} + 12}
\newcommand{\oneQotocFour}{64 \cos^{4}{\left(4 \omega t \right)} - 128 \cos^{2}{\left(4 \omega t \right)} + 64}

\newcommand{\twoQotocTwo}{256 - 256 \cos{\left(4 h t \right)}}
\newcommand{\twoQotocThree}{2 \left(256 - 256 \cos{\left(4 h t \right)}\right) \left(256 \cos{\left(4 h t \right)} + 256\right) + \left(256 \cos{\left(4 h t \right)} + 256\right)^{2}}
\newcommand{\twoQotocFour}{4 \left(256 - 256 \cos{\left(4 h t \right)}\right)^{2} \left(256 \cos{\left(4 h t \right)} + 256\right)^{2}}

\begin{document}

\title[Higher-order operator correlators]{Quantum information spreading via higher-order operator correlators}


\author*[1,2]{\fnm{M} \sur{S\"uzen}}\email{mehmet.suzen@physics.org}
\affil*[1]{\orgdiv{Resident Scientist}, \orgaddress{ \city{Assia}, \postcode{CY 5561}, \country{Cyprus}}}
\affil*[2]{\orgdiv{American Physical Society, Member}, \orgaddress{\city{College Park}, \postcode{MD}, \country{United States}}}

% \affil*[1]{\orgdiv{Department}, \orgname{Organization}, \orgaddress{\street{Street}, \city{City}, \postcode{100190}, \state{State}, \country{Country}}}
% \affil[2]{\orgdiv{Department}, \orgname{Organization}, \orgaddress{\street{Street}, \city{City}, \postcode{10587}, \state{State}, \country{Country}}}

% \affil[3]{\orgdiv{Department}, \orgname{Organization}, \orgaddress{\street{Street}, \city{City}, \postcode{610101}, \state{State}, \country{Country}}}


\abstract{
We provide an exact analytical benchmarks into measuring the quantum information spreading by 
introducing N-operator correlators. Leveraging the idea from N-point 
correlation function from cosmology, we formulate a generalized 
Out-of-Time-Ordered Correlator (OTOC), called N-Operator 
Correlator (NOC). Such formulation can reduce the effect of 
restricted analysis associated with selecting only two operators 
for OTOC. NOC can take into account a larger set of measurement 
probes in measuring the degree of quantum information spreading. 
We demonstrate these concepts with two analytically tractable 
systems with computing thermal averages of operator time 
evolutions at a finite temperature: a single qubit in longitudinal 
field and isotropic XXX model with a transverse field, two-qubit system. 
Our analytical benchmark serves as a basic introduction to generalized operator 
correlators in quantum mechanics.
}


%\pacs[MSC Classification]{35A01, 65L10, 65L12, 65L20, 65L70}

\maketitle

\section{Introduction}

The question of how and when unitary evolution of quantum states 
forgets their initial state, is currently an open research 
area \cite{swingle18a, xuswingle24}, in the settings of condensed matter 
and gravitational systems, prominently via characterizing the 
scrambling of quantum information \cite{hayden07, sekino08,
iyoda18sq,susskind20}. The primary notion of scrambling is tracking the 
initial state until it reaches to thermalization, checking if 
initial state is indistinguishable from the starting state by any 
local probes. This is closely related to quantum Lyapunov stability and 
butterfly effect \cite{peres84, shenker14bhbe, 
maldacena16, cotler17, rozenbaum17, campisi17, xu20, 
parker19, bala22, suzen2026s}, 
along with quantum chaos for a statistical measures. 
\cite{bohigas84, berry84, berry89}. 

Out-of-Time-Ordered Correlator (OTOC) \cite{kitaev15, hashimoto17, gukitaev22} 
express how information is spreading for a quantum mechanical system over 
time, given two other probes (operators), originating from measuring 
operator divergences in superconductive systems
\cite{larkin69}. Experimental realizations 
are quite challenging \cite{swingle16m, green22}, initial suggestions 
with NMR based quantum simulators were discussed \cite{li17m} and quite
recently direct measurement were realized \cite{googleq25, liang25, fricke26}. 
In this context, connections to irreversibility are also studied 
prominently by Zurek's team \cite{yanzurek20a}. 

In the Section \ref{sec:otoc}, we give foundational introduction 
and analytical computation details within the quantum dynamics setting. 
We provide two analytically tractable systems as an example. 
Generalization to N-operators are established in the Section 
\ref{sec:noc}. We use the same two systems in computing three 
and four operator correlations. We conclude our study in
the Section \ref{con}.

% Tutorial of Xu-Swingle: that's super heavy-not 
%                         so pedagogical.
% Concept: Density matrix and 
%          reduced density matrix, tracing out.

\section{Out-of-Time-Ordered Correlators in canonical ensemble} \label{sec:otoc}

We investigate the behavior of a quantum dynamical system in the finite 
temperature. Let's $H$ represents the Hamiltonian of the system and given two
other probes $P_{1}(t)$ and $P_{2}(t)$, information spreading 
$C(t) \equiv C(P_{1},P_{2},t)$ is measured via a 
commutator $\mathscr{O}(t)$ over thermal averages, in a 
quantum canonical ensemble \cite{pathria},

\begin{eqnarray}  
C(t)           & = &  \langle \mathscr{O}(t) \rangle_{T}  \\
\mathscr{O}(t) & = & [P_{1}(t),P_{2}(0)]^{\dagger}
                     [P_{1}(t),P_{2}(0)].
\end{eqnarray}  
The time evolution of the probes $P$ can be computed via 
Heisenberg unitary evolution \cite{sakurai20},
\begin{equation}
P_{1}(t) = \exp(iHt) P_{1}(0) \exp(-iHt).
\end{equation}
This kind of unitary evolution in $C(t)$ as a square commutation 
relation measures the {\it quantum information spreading} as we 
probe the forward and backward evolution with respect to 
the initial condition. This is the essence of the term 
"Out-of-Time-Ordered", and correlation comes from the fact 
that we tract commutation relation as a measure of information 
scrambling.

The thermal averages follows 
quantum canonical ensemble at a fixed temperature $T$, 
where inverse temperature is defined by 
$\beta=1.0/k_{B}T$, $k_{B}$ being Boltzmann constant. 
The density matrix $\rho$ in the quantum canonical ensemble
at inverse temperature $\beta$, can be defined as follows.
First, we find the eigenvalue and eigenvector pairs
$(E_{k}, v_{k})$ for the Hamiltonian $H$, $k$ being up to
a matrix order. $\rho$ reads,
\begin{equation}
\rho(\beta) = V Q V^{\dagger}.
\end{equation}
$V$ is the matrix formed by eigenvectors.
We define a real valued binary indicator row vector $\langle I_{k}|$ where 
$\langle I_{k}|$'s $k^{th}$ element is one and the rest 
of the vector is zero. Using this, we use a formulation 
to show mathematical structure of stacking eigenvectors 
sequentially,
\begin{equation}
V = \sum_{k} \langle I_{k}|  \otimes | v_{k} \rangle.
\end{equation}
Thermal matrix $Q$ is formed by the partition function, 
which is the normalization 
constant, $Z(\beta) =  \sum_{k} \exp(-\beta E_{k})$
and expected Boltzmann probabilities $p_{k} = exp(-\beta E_{k})$.
The normalized probabilities $\bar{p}_{k} =Z^{-1}(\beta) p_{k}$,
hence the thermal matrix reads $Q(\beta) = diag(\bar{p}_{k})$. \\
Consequently, the thermal averaging at finite temperature
reads,
\begin{equation}
C(P_{1}, P_{2}, t, ;\beta) = Tr(\rho(\beta) \mathscr{O}(t)).
\end{equation}
This measures the {\it quantum information spreading} given probes $P_{1}$, $P_{2}$.
This is not a measure symmetric, as one of the probe is fixed and the other evolves 
over time.

Computing such correlators require solving an eigenvalue problem 
for the Hamiltonian $H$, forming the density matrix at a 
fixed $\beta$ and evolve the chosen two probing operators 
$P_{1}(0)$ and $P_{2}(0)$ over time via Heisenberg unitary operator 
evolution. We also need to use the spectral decomposition \cite{arfken7} 
for computing $\exp(Hit)$ and $\exp(-Hit)$ with {\it orthonormal eigenvectors}. 
They read as follows,
\begin{eqnarray}
\exp(iHt) & = &  V R_{+} V^{\dagger} \\
\exp(-iHt) & = &  V R_{-}  V^{\dagger}.
\end{eqnarray}
The matrix $R_{\pm}$ is formed with the eigenenergies as follows,
\begin{eqnarray}
rp & = & [\exp(it E_{1}), ..., \exp(it E_{k}))] \\
rn & = & [\exp(-it E_{1}), ..., \exp(-it E_{k}))]
\end{eqnarray} 
placing these vectors into diagonals, $R_{+}=diag(rp)$ and 
$R_{-}=diag(rn)$. We 
apply this framework for two simple problems, single qubit 
in a longitudinal field and the two-qubit unitary system obeying
Hilbert space composition rules \cite{von96}. 

\begin{figure}[ht!]
\centering
\subfloat[\label{q1otoc2}]{\includegraphics[width=0.33\textwidth]{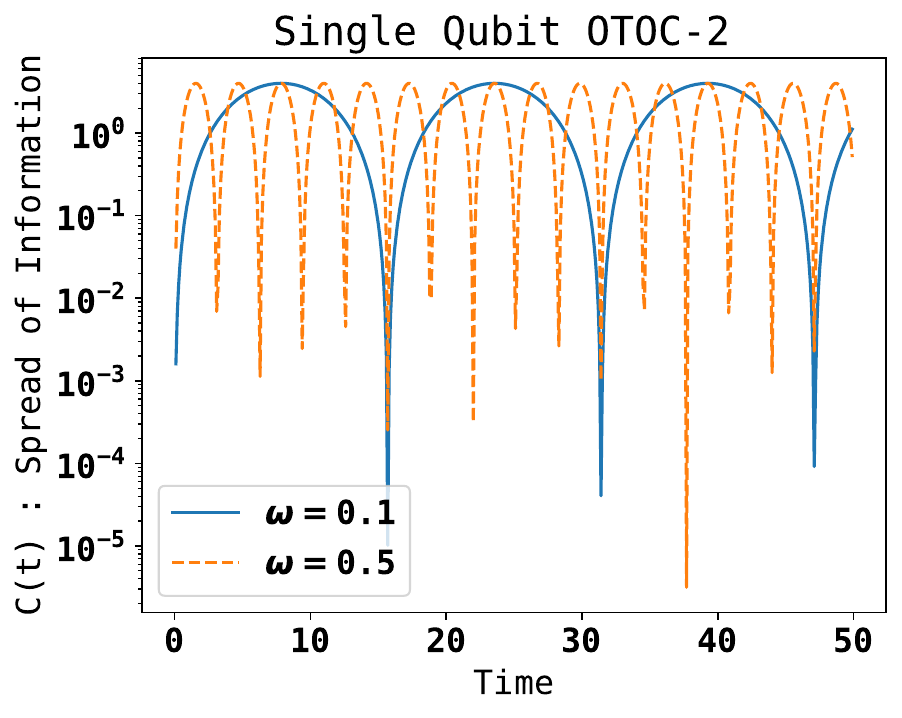}} 
\subfloat[\label{q1oto3}]{\includegraphics[width=0.34\textwidth]{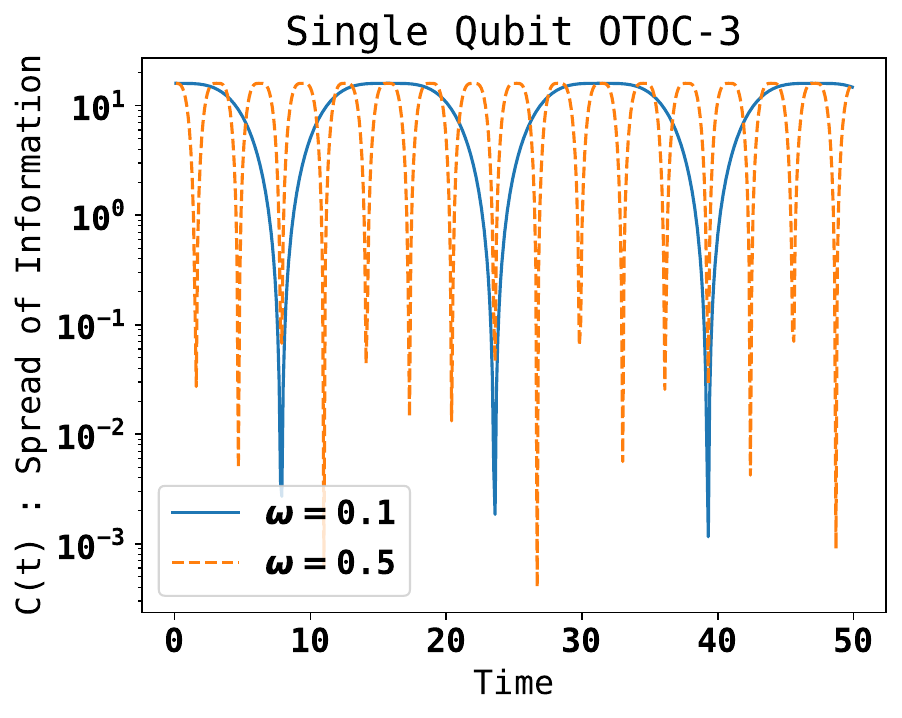}} 
\subfloat[\label{q1otoc4}]{\includegraphics[width=0.33\textwidth]{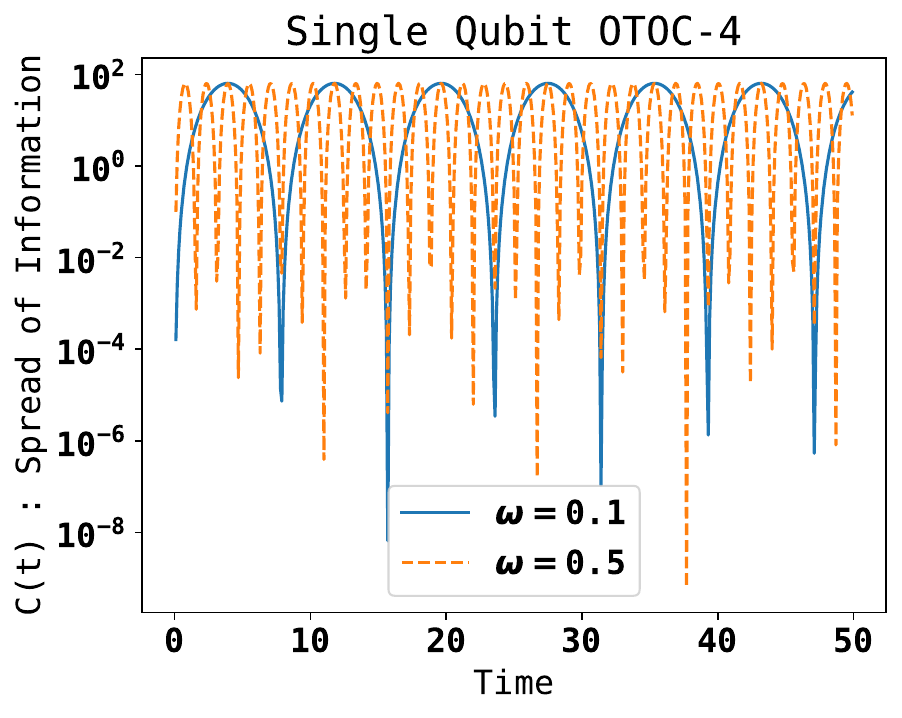}}
\caption{
$C(t)$ over time for the single qubit in traverse field: 
(a) OTOC-2, Standard OTOC with $P_{1}$ and $P_{2}$ operators.
(b) OTOC-3, $NOC_{3}$ with $P_{1}$, $P_{2}$ and $P_{3}$ operators.
(b) OTOC-4, $NOC_{4}$ with $P_{1}$, $P_{2}, P_{3}$ and $P_{4}$ operators.
}
\end{figure}

\subsection{Single Qubit Hamiltonian}

A fermion spin 1/2 particle qubit is considered in a longitudinal 
field with the strength $\omega$, then the Hamiltonian reads,
\begin{equation}
H = \omega \sigma_{z}.
\end{equation}
We choose two operators $P_{1}(0)=P_{2}(0)=\sigma_{y}$
as a probe, so called YY probes. $\sigma_{z}$ and $\sigma_{y}$ are defined as
follows. 
\begin{equation}
\sigma_{z} = \begin{pmatrix} 1 & 0 \\ 0 & -1 \end{pmatrix}, \quad 
\sigma_{y} = \begin{pmatrix} 0 & -i \\ i & 0 \end{pmatrix}, \qquad
\sigma_{x} = \begin{pmatrix} 0 & 1 \\ 1 & 0 \end{pmatrix}.
\end{equation}
This is one the simplest system one can use. Some of the 
analytical results from the framework using YY probes, 
$R_{\pm}, P(t), F(t)$ and $C(t)$ are given as follows,

\begin{equation}
R_{+} = \oneQRp \quad
R_{-} = \oneQRm,
\end{equation}

\begin{equation}
C(t) =  \oneQotocTwo
\end{equation}
We plot $C(t)$ for standard OTOC in Figure \ref{q1otoc2}, under different field conditions. The
temperature dependence cancels. We see that
information spreading oscillates, not spreading completely but remain bounded.

\subsection{Two-Qubit XXX Model}

\begin{figure}[ht!]
\centering
\subfloat[\label{q2otoc2}]{\includegraphics[width=0.33\textwidth]{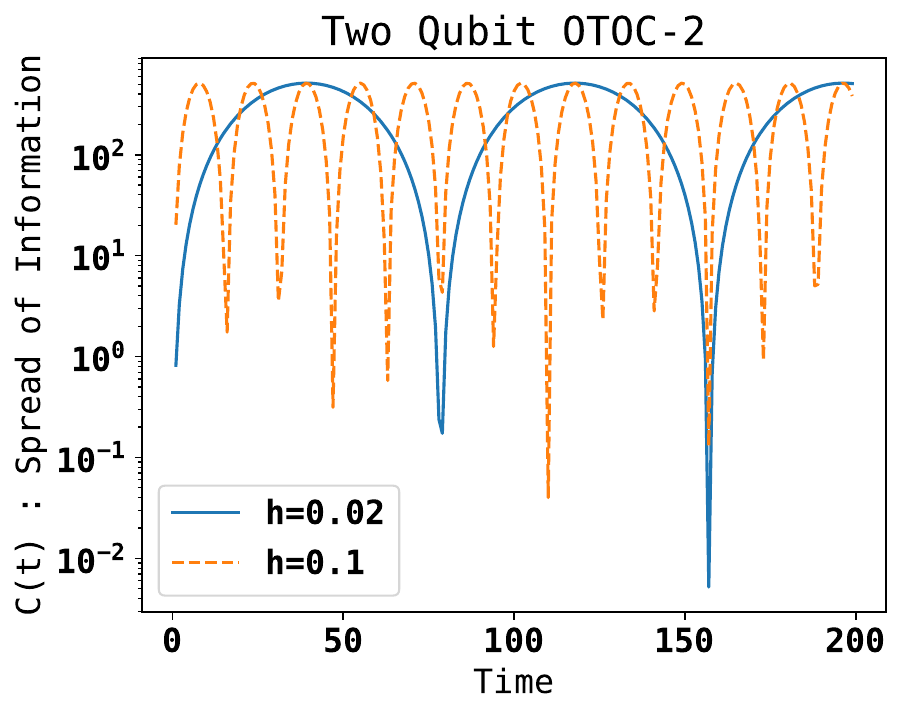}} 
\subfloat[\label{q2oto3}]{\includegraphics[width=0.33\textwidth]{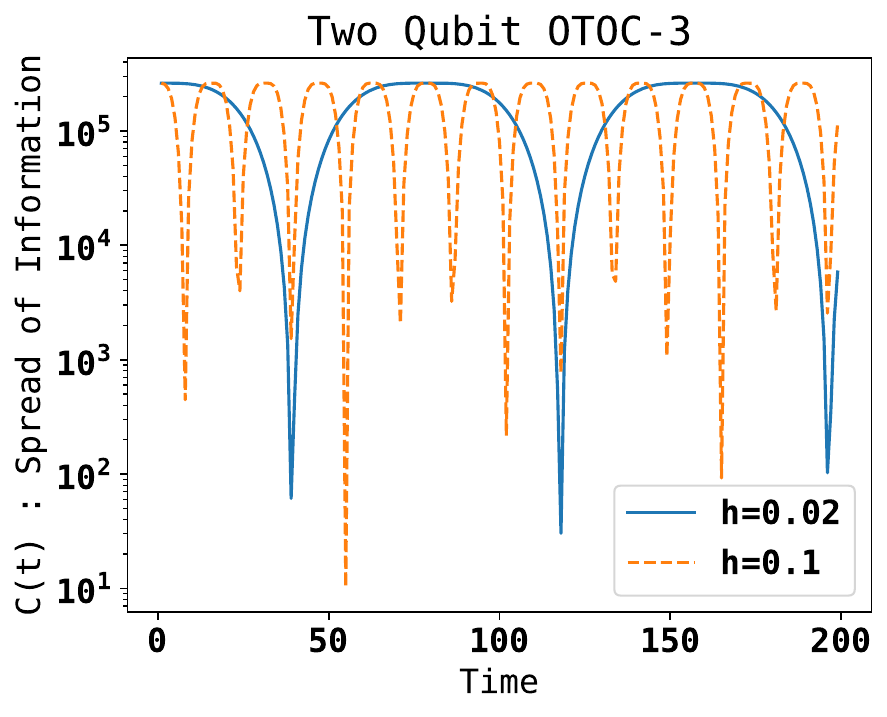}} 
\subfloat[\label{q2otoc4}]{\includegraphics[width=0.33\textwidth]{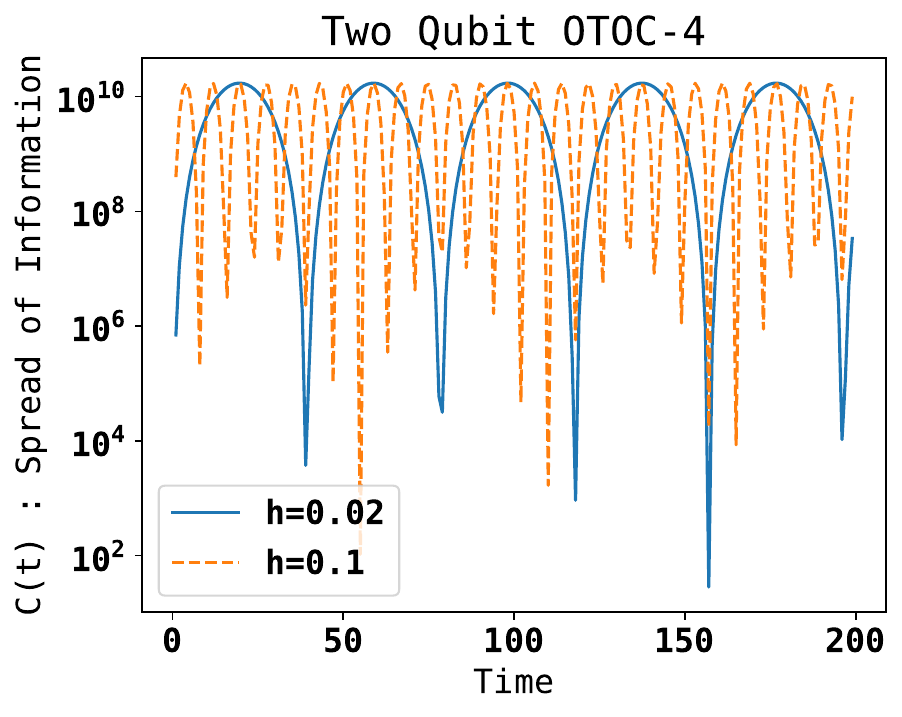}}
\caption{
$C(t)$ over time for the single qubit in traverse field: 
(a) OTOC-2, Standard OTOC with $P_{1}$ and $P_{2}$ operators.
(b) OTOC-3, $NOC_{3}$ with $P_{1}$, $P_{2}$ and $P_{3}$ operators.
(b) OTOC-4, $NOC_{4}$ with $P_{1}$, $P_{2}, P_{3}$ and $P_{4}$ operators.
}
\end{figure}

Quantum \cite{degennes63, wells14} and classical \cite{suzen14} spin chains 
are probably a gold standard systems in studying prominent physical phenomena, 
such as exhibiting diffusive power laws in ergodicity anomalies 
\cite{liang25, suzen26ps} and with the correlators \cite{lin17o}. Here, we 
define two-qubit system using Pauli operators, isotropic XXX model with a transverse field. 
Hamiltonian for two qubit system reads, nearest neighbor and external field interactions, 
\begin{equation}
H(s_{1}^{x,y,z},s_{2}^{x,y,z},J,h) = J \large(s_{1}^{x} s_{2}^{x} + s_{1}^{y} s_{2}^{y} 
+ s_{1}^{z} s_{2}^{z} \large) 
+ h \large( s_{1}^{x} + s_{2}^{x}\large). \nonumber
\end{equation}
The spin variables are expressed with Pauli operators, 
\begin{eqnarray}
s_{1}^{x,y,z} & = & \sigma_{x,y,z} \otimes \mathbb{I}_{2} \\
s_{2}^{x,y,z} & = & \mathbb{I}_{2} \otimes \sigma_{x,y,z}.
\end{eqnarray}
We recall that $\sigma_z = \begin{pmatrix} 1 & 0 \\ 0 & -1 \end{pmatrix}$, and 
identity matrix in the Pauli group $\mathbb{I}_{2} = \begin{pmatrix} 1 & 0 \\ 0 & 1 \end{pmatrix}$.
We choose the following probes, YY probes,
\begin{equation}
P_{1} = P_{2} = s_{1}^{y} + s_{2}^{y}. 
\end{equation}
The Hamiltonian reads, 
\begin{equation}
H = \left[ \begin{matrix} 
    J & h   & h   & 0 \\ 
    h & - J & 2 J & h \\ 
    h & 2 J & - J & h \\
    0 & h   & h   & J 
  \end{matrix} \right]. 
\end{equation}
We plot $C(t)$ for two qubit system in Figure \ref{q2otoc2} under different external field 
conditions,
\begin{equation}
C(t) = \twoQotocTwo
\end{equation}
We observe a cyclic spreading with higher frequency compare to single qubit, indicating 
stronger coupling.

\section{N-Operator Correlators} \label{sec:noc}

The idea of correlation originates from physics and advanced with 
the research in statistics \cite{wright21}. Extensively used in
all areas of sciences, defined for two variates (points). Extension
to N-points is prominently occurred in cosmological context due to 
complexity in large scale structure \cite{desjacques, takada03} associated with 
the matter density $\rho$. The problem of how to express such N-correlators are 
resolved with hierarchical approach, by applying correlation in a cyclic
manner for $3-point$ correlation functions, denoted by 
3PCF \cite{takada03, aviles, sugiyama} reads, via combining pairwise correlations, 
\begin{eqnarray}
\rho(x_{1}, x_{2}, x_{3}) & = & \rho(x_{1}, x_{2}) \rho(x_{3}, x_{1}) + \\
                          & & \rho(x_{2}, x_{3}) \rho(x_{3}, x_{1}) + \nonumber \\ 
                          & & \rho(x_{3}, x_{1}) \rho(x_{1}, x_{2}), \nonumber
\end{eqnarray}
as the simplest example $N > 2$ using three variates.
We use this idea for N-Operator Correlators (NOCs) as follows. Denoting 
$N$ probing operators with $P_{k}(t)$ that $k=1,..N$ and $C_{ij}=C(P_{i}, P_{j},t)$ 
the correlation in between 
$i^{th}$ and $j^{th}$ probes. The {\it N-operator Correlators (NOCs)} reads,
\begin{equation}
\label{nocPi}
C_{N}(P_{1}, .., P_{N},t) = \sum_{i=1}^{N-1} \left[ \prod_{j=i}^{i+N-2} C_{j,j+1} \right],
\end{equation}
where the indices are taken cyclically ($P_{N+k} \equiv P_k$), up to a single cycle only.
In this definition weights from different pair-wise correlations are taken equally. 
In more applied cases such as in simulations of correlations functions \cite{thirumalai}, 
in our formulation of $C_{N}(P_{1}, .., P_{N},t)$ weights can be assigned in the 
sum-product formula.

In our examples we use $N=3$ and $N=4$ operators, then N-Operator Correlations 
in these cases read,
\begin{eqnarray}
C_{3}(P_{1}, P_{2}, P_{3},t) & = & C(P_{1}, P_{2}, t) C(P_{2}, P_{3},t) + C(P_{2}, P_{3}, t) C(P_{3}, P_{1},t) + \nonumber \\
                             &   & C(P_{3}, P_{1}, t) C(P_{2}, P_{3},t), \\
C_{4}(P_{1}, P_{2}, P_{3}, P_{4},t) & = & C(P_{1}, P_{2}, t) C(P_{2}, P_{3},t) C(P_{3}, P_{4},t) C(P_{4}, P_{1},t) +  \nonumber \\ 
                                    &   & C(P_{2}, P_{3}, t) C(P_{3}, P_{4},t) C(P_{4}, P_{1},t) C(P_{1}, P_{2},t) +  \nonumber \\ 
                                    &   & C(P_{3}, P_{4}, t) C(P_{4}, P_{1},t) C(P_{1}, P_{1},t) C(P_{2}, P_{3},t) +  \nonumber \\ 
                                    &   & C(P_{4}, P_{1}, t) C(P_{1}, P_{2},t) C(P_{2}, P_{3},t) C(P_{3}, P_{4},t).
\end{eqnarray}

There is a slight terminological difference in between quantum and
classical cosmological context. The count of points in cosmology are 
distinct vector variables, we use instead operators. In the quantum
literature, usually terminology of {\it four-point correlation} is 
used as a mathematical object, though only two operators appear with 
different time origins in the formulations. This nuance appears due to 
time-correlation nature in the quantum setting. We follow 
cosmological approach in counting operators, rather than time origins. 
However, this is just a convention we used, doesn't change the underlying 
physical generalization. 

The choice of different probes should be guided by available observables,
physical insights and experimentally accessible operators. Similarly,
any mathematical recipe to choose different operators, should follow
physical properties. In our examples, we choose from 
available operators in our simple systems, and they aren't unique 
choices. 

In our definition, we actually use quantum information spreading $C(t)$.
This is apparent in our notation $C_{N}(P_{1}, .., P_{N},t)$. 
The same cyclic combination approach can also be applied into any kind of quantum 
correlations \cite{adesso16} in a pairwise settings in building 
multi-dimensional correlations. We concentrate on OTOCs, as
time-correlation function in quantum 
mechanics \cite{yanzurek20a, alhambra20, tokmakoff25}.

\subsection{Single Qubit Hamiltonian with N-probes}

An initial step in generating $N=3,4$ correlators, $C_{3}(P_{1}, P_{2}, P_{3},t)$ 
and $C_{4}(P_{1}, P_{2}, P_{3},P_{4},t)$, for a single qubit, we choose the following
probing operators $P_{1}=P_{2}=P_{4}=\sigma_{y}$ and $P_{3}=\sigma_{x}$. Computing 
the correlators analytically we reach to the following expressions for NOCs,   
\begin{eqnarray}
C_{3}(P_{1}, P_{2}, P_{3},t) & = & \oneQotocThree \nonumber \\
C_{4}(P_{1}, P_{2}, P_{3},P_{4},t) & = & \oneQotocFour .
\end{eqnarray}

We observe higher oscillation wavelengths at higher-order correlators as seen
in Figure \ref{q1oto3} and \ref{q1oto4} and. This indicates using multiple probes 
provide a rich behavior. For example, for $C_{4}$, we observe 
stronger depth when spreading retreats, before it spread again.    

\subsection{Two-Qubit XXX Model with N-probes}

In case of our XXX model with traverse field, we choose the following probes, 
\begin{equation}
P_{1} = P_{2} = P_{4} = s_{1}^{z} + s_{2}^{z} \quad P_{3} = s_{1}^{x} + s_{2}^{x}.
\end{equation}

Using the canonical ensemble framework, we analytically find the exact solutions,
at $h=0.02, 0.1$,

\begin{eqnarray}
C_{3}(P_{1}, P_{2}, P_{3},t) & = &  \twoQotocThree \nonumber \\
C_{4}(P_{1}, P_{2}, P_{3},P_{4},t) & = & \twoQotocFour  .
\end{eqnarray}

In the case of two-bit isotropic XXX model with a transverse field, in 
Figure \ref{q2oto3} and \ref{q2oto4}, we observe richer variation in 
higher-order cases. In the case of $C_{4}$ oscillations are more pronounced 
compare to $C_{3}$. Temperature and local strengths are cancels out for this 
operator selections. We particularly choose this for a simpler spread profiles.
Temperature and local field dependence would be observed with different 
operator selections, leading to much more complex analytical solutions.

\subsection{Higher-order and higher-point functions}

We need to differentiate in between higher-order and higher-point OTOCs. 
The {\it higher-order} implies multiple probes using the two time origins.
On the other hand, {\it higher-point} implies using standard two 
probe approach but using many different time origins to tract them. 
Both approaches enriches the measurement of quantum information spreading.

There are alterations of OTOC definition to accommodate the higher-order and higher-point 
approach. Alterations place all additional probes or time-origins within 
the same correlation function, such as $\langle O_{i}(0) O_{j}(t) \rangle$ or 
$\langle O_{1}(t_{k}) O_{2}(t_{l}) \rangle$, where $i,j$ are multiple probes 
in two time origins and $k,l$ are multiple time origins for two probes.
Such as 2N-point OTOC in SUSY quantum mechanics \cite{das24}, 6-point OTOC for 
describing field theories \cite{haelh18}, 4-point OTOC for entangled 
qubits \cite{garcia21}, n-point OTOC for random matrices \cite{jahnke25fp} 
and higher-order OTOCs in measuring correlation in superconducting circuits 
\cite{googleq25}.

Our cyclic formulation presented for N-operator correlation in Equation 
\ref{nocPi} represents a distinct approach in using multiple probes. 
The formulation distinctly tracks standard OTOC in pair-wise manner 
without altering the primary definition of OTOC. This is physically more 
robust as a representation of physical interactions.

\section{Conclusion} \label{con}

Inspired `from cosmology, we have generalized the concept of OTOCs,
building correlations based on N-operator probes, so-called NOCs,
more precisely, {\it Cyclic N-Operator Correlators} (CNOCs).
Such NOCs will enable us to use more measurements simultaneously
in quantifying overall information spreading in quantum systems. 
Experimentally, this can be interesting if multiple components are 
interacting and multiple probes are used in quantum control systems
in tracking the global correlation. Moreover, as we have demonstrated, 
theoretically, NOCs can lead to richer set of spread profiles, providing 
more insight on the physics of multiple probes. NOC is a complimentary tool to OTOCs.
It could help in choosing a more representative or physically
most relevant operator pairs for OTOCs. \\

% \ack{We thank Y.Suzen for her kind support of 
% this work within the quantum dynamics project.}

{\it Data availibility:}
Supporting symbolic workflow is available on request, here \cite{suzen26nocsup}.

{\it Acknowledgments:} Y. S\"uzen for her kind support and encouragements.
We thank Rathindra Nath Das (MIT) for sharing his work on multi-time origin 
correlations within SUSY quantum mechanics. 
\nocite{*}
\bibliography{suzen}

\end{document}